2015

# The Quantitative Reasoning for College Science (QuaRCS) Assessment, 1: Development and Validation

Katherine B. Follette
*Kavli Institute of Particle Astrophysics and Cosmology, Stanford University*, kfollette@stanford.edu

Donald W. McCarthy
*University of Arizona*

Erin Dokter
*University of Arizona*

Sanlyn Buxner
*University of Arizona*

Edward Prather
*University of Arizona*

Follow this and additional works at: http://scholarcommons.usf.edu/numeracy

Part of the Educational Assessment, Evaluation, and Research Commons, and the Science and Mathematics Education Commons

# The Quantitative Reasoning for College Science (QuaRCS) Assessment, 1: Development and Validation

**Abstract**

Science is an inherently quantitative endeavor, and general education science courses are taken by a majority of college students. As such, they are a powerful venue for advancing students' skills and attitudes toward mathematics. This article reports on the development and validation of the Quantitative Reasoning for College Science (QuaRCS) Assessment, a numeracy assessment instrument designed for college-level general education science students. It has been administered to more than four thousand students over eight semesters of refinement. We show that the QuaRCS is able to distinguish varying levels of quantitative literacy and present performance statistics for both individual items and the instrument as a whole. Responses from a survey of forty-eight Astronomy and Mathematics educators show that these two groups share views regarding which quantitative skills are most important in the contexts of science literacy and educated citizenship, and the skills assessed with the QuaRCS are drawn from these rankings. The fully-developed QuaRCS assessment was administered to nearly two thousand students in nineteen general education science courses and one STEM major course in early 2015, and results reveal that the instrument is valid for both populations.

**Cover Page Footnote**

**Kate Follette** is a postdoctoral researcher at the Kavli Institute for Particle Astrophysics and Cosmology at Stanford University. The majority of this work was conducted at the University of Arizona, where she was supported by an NSF Graduate Research Fellowship and an NSF Transforming Undergraduate Education in STEM grant. She taught general education astronomy courses as an adjunct instructor at Pima Community College from 2009-2014. Her science research focuses on the discovery and characterization of extrasolar planets and the disks of gas and dust from which they form.

**Don McCarthy** is a research astronomer and Distinguished Outreach Professor at The University of Arizona where he teaches science to undergraduates in the general education program and also astronomy courses for undergraduate astrophysics majors and minors. His research expertise in infrared astronomy focuses on the development and application of techniques for high angular resolution from large ground- and spaced-based telescopes with an emphasis on exoplanets and Solar System objects. He received the 2012 Education Prize from the American Astronomical Society.

**Erin Dokter** is an associate professor of practice in the Office of Instruction and Assessment at the University of Arizona, where she serves as the Coordinator for the Certificate in College Teaching program, and as adjunct faculty in the Agricultural Education department. She has taught general education astronomy courses as adjunct faculty at San Diego State University, Grossmont College, Southwestern College, and Pima Community College. Her research and teaching interests revolve around educational development, cross-disciplinary learner-centered pedagogies, and active learning spaces.

**Sanlyn Buxner** is an assistant research professor in the department of Teaching, Learning and Sociocultural Studies at the University of Arizona. She teaches introductory science and research methods courses and supports institutional assessment of graduate and undergraduate programs. Her research includes examining science literacy and quantitative literacy in undergraduate science students and studying the impact of research and industry experiences for K12 teachers on classroom practice and student outcomes.

**Edward Prather** is an Associate Professor in the Department of Astronomy and Steward Observatory at the University of Arizona. He serves as the founder and Executive Director of the NASA-funded Center for Astronomy Education. Ed's scholarship focuses on college-level STEM education research and faculty professional development.

# Introduction

Missing in much of the literature on numeracy[1] are specific, research-validated strategies for mitigating innumeracy and improving the quantitative skills of American students and adults. Development of such strategies is fundamentally difficult, requiring both evidence-based instructional techniques and assessments capable of measuring learners' quantitative abilities. In this article, we present one such assessment specifically targeted toward students in general education[2] college science courses – the Quantitative Reasoning for College Science (QuaRCS) assessment.

In the next section, we describe motivations for developing the QuaRCS and outline the questions that we set out to address in designing the instrument. We explain why we focus on general education college science courses rather than directly targeting mathematics or numeracy courses, and we draw on statistics about the demographics of our sample to support our claims. In the third section, we describe the results of a survey of science and mathematics educators that led to the selection of specific quantitative skills for assessment. Next, we provide an overview of the instrument itself, including the format and types of questions. In the penultimate section, we discuss a Classical Test Theory analysis of the instrument, and we close with a summary of our findings.

# The Need for a Numeracy Assessment Instrument for General Education Science Courses

The academic literature contains a plethora of compelling arguments for why numerical skills are critical to success in everything from managing one's personal health (Schwartz et al. 1997, Apter et al. 2009, Brown et al. 2011) to understanding the informed consent process (Couper and Singer 2009). As research has shown, the relevance of numeracy extends to other, potentially life-altering events, such as one's ability to access opportunities for employment (Kirsch et al. 1993, Charette and Meng 1998) and make sound financial decisions (Gerardi et al. 2013). Unfortunately, there remains much to be done to support American adults in achieving quantitative literacy (e.g., Lemke and Gonzales 2006, Goodman et al. 2013). The issue of rampant innumeracy has gained such attention in recent years that it frequently appears in national news headlines (e.g., Perez-Pena 2013, Green 2014) and has been the subject of numerous works of popular literature (e.g., Seife 2010, Schneps and Colmez 2013, Bennett 2014). As the importance of numerical skills for success in modern life has been argued in many other venues, we will not belabor it here.

---

[1] We will use this term as well as quantitative literacy and quantitative reasoning interchangeably throughout this paper.
[2] Courses fulfilling University-level general education or distribution requirements.

Even among college or college-bound students with strong academic preparation, quantitative literacy falls below expectations, necessitating developmental or remedial coursework (e.g., Lee 2012, California State University Proficiency Rates n.d., Bettinger et al. 2013). For example, among college-bound students who took the SAT in 2014, the mean score on the mathematics portion of the assessment was independent of the number of mathematics course taken (College Board 2014). Results like these have prompted calls to action, including the promotion of numeracy education across the higher education curriculum (e.g., Steen 1999, 2001, 2004, Lutsky 2008, Hillyard 2012, Elrod 2014) and measurement of student experiences with mathematics in their college courses (e.g., Dumford and Rocconi 2015).

With the goal of measuring changes in quantitative abilities, several quantitative literacy assessment instruments have already been developed.[3] However, many are proprietary (e.g., Hollins University, University of Akron, University of Virginia, Norfolk State University), costly (e.g., ACT WorkKeys, the Graduate Record Examination, Insight's Test of Everyday Reasoning - Numeracy, James Madison University's Quantitative Reasoning Test), or focused on specific numeracy domains (generally statistics, e.g., e-ATLAS, Levels of Conceptual Understanding in Statistics (LOCUS), and Milo Shield's Statistical Literacy Tests). Several others are in development, but have yet to be rigorously tested for validity and reliability. The most rigorously validated non-proprietary general-purpose numeracy assessment instrument developed to date is the Quantitative Literacy and Reasoning Assessment (QLRA, Gaze et al. 2014). Like the QuaRCS, the QLRA assesses both attitudes and numerical skills, and we believe the two assessments are highly complementary.

While the QLRA took a "top down" approach by combining several already existing numeracy instruments and relying heavily on the considerable expertise of its developers for content selection, we took a "bottom up" approach, surveying science and numeracy educators in order to focus on the skills most relevant for science literacy. The QuaRCS also contains a more expansive bank of attitude and academic background questions, and the post-semester instrument includes a series of questions about the course in which it is administered. These differences were driven by the specific questions that we set out to answer in developing the QuaRCS, as laid out in the following.

---

[3] A comprehensive list of available numeracy assessments is available at serc.carleton.edu/NICHE/ex_qr_assessment.html

## *Why General Education Science Courses?*

Numerous studies have identified general education and/or introductory science courses as venues with the potential to complement the development of numerical skills (e.g., Powell and Leveson 2004, Bray Speth et al. 2010, Hester et al. 2014, Hathcoat et al. 2015). For several decades now, much of the dialogue surrounding the teaching of general education introductory science courses has been focused on instilling "science literacy" (Rutherford and Ahlgren 1991, DeBoer 2000).[4] Although quantitative literacy is only rarely explicitly considered in this context (e.g., Meisels 2010), the skills that comprise science literacy, such as the ability to interpret and understand scientific information in the media, are often quantitative in nature. Therefore, investigating the role that general education science instructors can play in instilling quantitative literacy is crucially important to the future of science literacy and the role of quantitative literacy within it.

Our emphasis on general education courses, taken predominantly by non-STEM majors, was motivated by a suspicion that, not only are these courses often the final science course of a formal education, they are also perceived by students to be their final *quantitative* course. In our Spring 2015 cohort of 1480 general education science students at a large research-intensive University in the Southwestern United States, 55% intended to take only the University-required number of science courses (2). When asked how many math courses they had taken or intend to take in college, 10% ($N=147$) reported that they did not intend to take any math courses, and 23% ($N=337$) intended to take just one mathematics course. Furthermore, 68% ($N=1010$) of students indicated that they were enrolled in their general education science course "in order to fulfill a University requirement," and only 17% ($N=258$) were enrolled to fulfill a prerequisite for their major.

Even if students do take more quantitative courses than they anticipate, an emphasis on numerical skills in general education college science courses may still serve them well. In particular, since the majority of students in these courses are Freshmen or Sophomores (~40% Freshmen and ~30% Sophomores in our sample), a quantitative emphasis has the potential to set them on a path toward making quantitative reasoning a part of their "academic toolbox" for the remainder of their college career.

Many universities and colleges require that their students take at least one science course before graduating, and these courses are therefore taken by a broad cross-section of the college population. Indeed, we find that the reported majors of students in our sample match the statistics of awarded bachelor's degrees for the University to within 5%. The one exception is STEM majors, who are less likely to enroll in general education science courses. In fact, in our Spring 2015 general education cohort, only 8% ($N$=123) reported that they would major in science and 4% ($N$=64) in engineering, math or computer science, while these groups make up 37% of awarded degrees at the University.

---

[4] Arguments for why numerical skills are critical to science literacy have been discussed elsewhere (e.g., Follette and McCarthy 2012a, Follette and McCarthy 2012b, Follette and McCarthy 2014, McCarthy and Follette 2013, Rutherford and Ahlgren 1991).

Although other disciplines, particularly business and social sciences (26% and 25% of our population respectively), certainly can and should emphasize numerical reasoning skills, students are unlikely to perceive non-STEM disciplines as mathematical. We present evidence for this claim in Table 1 below, which shows the number of students who reported choosing their major "to avoid writing as much as possible" or "to avoid math as much as possible" among the entire population, and among several key majors. These choices were embedded among a number of other options for why students chose their major (the entirety of which is reported in Paper 2[5] of this series). Students choosing one of these two "avoidance" options represent just 17% of the entire cohort, but the relative statistics are both significant ($p$<0.05) and telling, as revealed in Table 1.

**Table 1.**
**Reason for Choosing Major by Major.**

|  | "To Avoid Writing" | "To Avoid Math" |
|---|---|---|
| All Students ($N$=1480) | 65 (4%) | 180 (12%) |
| STEM Majors ($N$=182) | 19 (10%) | 6 (3%) |
| Non-STEM Majors ($N$=1298) | 46 (4%) | 174 (13%) |
| Business Majors ($N$=382) | 16 (4%) | 15 (4%) |
| Social Sciences ($N$=376) | 7 (2%) | 66 (18%) |

In general, students are approximately three times as likely to report choosing a major "to avoid math" as they are to choose one "to avoid writing." Whereas STEM students are roughly three times as likely to report that they chose their major to avoid writing, these statistics are reversed among non-STEM majors, who are about three times as likely to choose their major to avoid math. Among social science majors, this trend is amplified even further, with students nearly 10 times as likely to report choosing their major to avoid math. Business majors are the only group in which the prevalence of both choices is below 5% of the population.

Additionally, 6% ($N$=93) of the students in our sample are future educators (education majors), and national studies of general education science courses suggest that this proportion is often much higher (as high as 40%, Lawrenz et al. 2005). This population is another important one to reach, as poor attitudes of teachers toward mathematics have the potential to carry over to their students and perpetuate the problem of innumeracy (Jackson and Leffingwell 1999, Beilock et al. 2010).

---

[5] Follette et al. (submitted)

## *Why Aren't These Courses Already Advancing Quantitative Skills?*

Despite the natural disciplinary entwinement of science and mathematics, instructors often shy away from a numerical emphasis in general education science courses, which are taken predominantly by non-STEM majors. This aversion is due at least in part to the well-documented innumeracy of American adults,[6] but we believe that the problem is more nuanced than a simple lack of skill; we believe it has much to do with student attitudes and expectations about these courses. In our experience leading workshops for science educators, there are many varieties of concern and resistance to the notion of incorporating quantitative skills into science courses, and science educators often report trying and failing. In workshop evaluations, educators report having been "dissuaded by the griping," "apologizing for the math," and "worrying about student evaluations." As we began to develop the QuaRCS, we focused on addressing three specific arguments against incorporating numerical skills into general education science courses.

*Argument 1: It's too late for them.* College science educators often express frustration at the numerical deficits of their students and the effect that this lack of skill has on their curricula. They also, however, voice the belief that if a student's previous education fails to instill the necessary skills to understand quantitative science, then it's unlikely that a single college course will correct this deficiency. Whether this concern is warranted remains to be seen, and so we designed the QuaRCs as a pre- *and* post-semester assessment in order to address it. By administering the QuaRCS at the beginning and end of a semester of science instruction at a variety of institutions across the country, we hope to begin to investigate whether college science educators can make a meaningful difference in students' quantitative abilities over the course of a single semester of instruction.

*Argument 2: My students are here to learn science, not math.* When we seek clarification of comments like this at our workshops, it appears that the concern is largely a question of balance between remediation of numerical skills and coverage of science content. The question of whether quantitative remediation

---

[6] This has been demonstrated in both national and international surveys including the National Assessment of Adult Literacy (Kutner et al. 2007), National Adult Literacy Survey (Kirsch et al.1993) and the Program for International Student Assessment (Organization for Economic Cooperation and Development 2012)

results in lower science content gains has been addressed in other studies, most notably Hester et al. (2014). That study showed that students in a Biology course that engaged in remediation of quantitative skills scored equally well on general Biology content questions post-semester as courses that didn't emphasize numerical reasoning. At the same time, the intervention course showed substantially higher gains on so-called "BioMath" questions (Biology questions housed in a quantitative context).

Given the work of Hester et al., we set out to answer a separate, though related, question with our assessment. Are students able to apply numerical skills when they encounter "real life" problems that require similar reasoning to problems that they might have encountered in a quantitatively rich science course? In other words, when quantitative skills are emphasized in context in a science classroom, are skill improvements ***transferable*** to other contexts?

***Argument 3: My good students will be bored.*** Although on average the quantitative skills of general education college science students are poor, the distribution is wide, as evidenced in Figure 1, which shows a histogram of scores for general education science students on the Spring, 2015 QuaRCS pre-semester assessment. This wide distribution of skills is a further concern of science educators, who worry that engaging in remediation will underserve or bore better-prepared students.

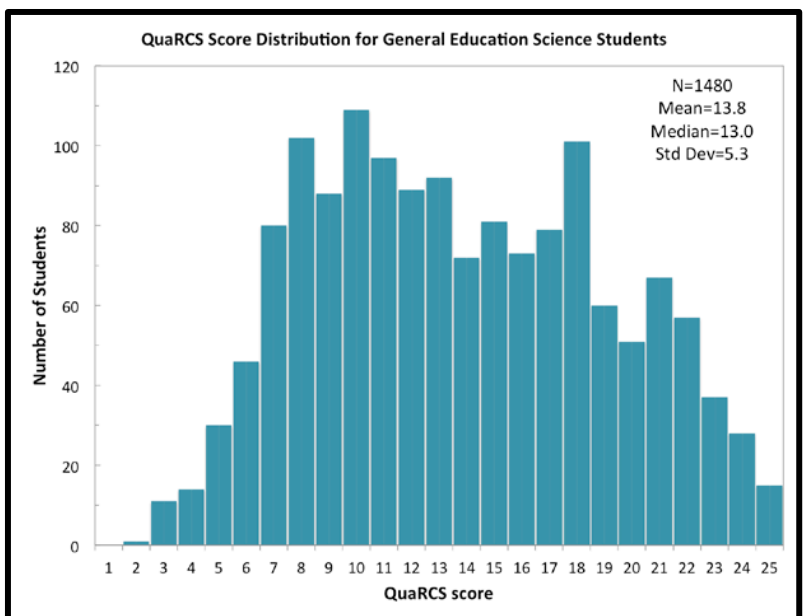

**Figure 1.** Distribution of scores (out of 25 total possible) on the Spring 2015 Pre-Semester Assessment. The range of student abilities is very wide, and the average ability level is low. Mean = 13.8 (55%), Median=13 (52%), Standard Deviation=5.3 (21%).

Although we suspect from previous studies of educational interventions in science disciplines (e.g., Rudolph et al. 2010) that a well-designed numeracy intervention will improve the skills of *all* students, we don't know of any existing literature supporting this claim in numeracy specifically. We will endeavor to understand this effect with the QuaRCS by analyzing gains among the students scoring highest on the pre-semester assessment, and by asking questions about how students feel about the *way* mathematics was emphasized in their course on the post-semester assessment.

## Skill Selection for the QuaRCS

There are numerous and conflicting opinions regarding what specific skills constitute numeracy; therefore, we began the development of the QuaRCS by selecting a subset of skills for assessment. In order to inform skill selection, we developed an educator survey asking respondents to rank numerical skills according to importance. The survey was intentionally populated with a mixture of (a) skills that appear frequently in the numeracy literature (e.g., proportional reasoning, arithmetic), (b) skills that are frequently invoked in the context of science literacy (e.g., graph reading, estimation), (c) "traditional" science skills whose utility in introductory courses is debated (e.g., scientific notation, algebra) and (d) advanced numerical skills that scientists use as a matter of course (e.g., calculus, exponents, logarithms). For each skill, a brief description was provided for calibration among educators.

The educator survey was pilot tested in paper form at two Astronomy education workshops in 2013 ($N$=34, 42). The question wording and choice of skills were modified based on analysis of these pilot surveys, including write-in responses for numerical skills missing from our original list. The final survey was administered online to several groups of University-level Astronomy ($N$=19) and Mathematics/Numeracy ($N$=29) educators. The list of skills and the statistics regarding their perceived importance in various contexts are shown in Figure 2.

On average, both groups of educators rank the majority of our skills above the midpoint of our ranking system ("important"), and there are just a handful of skills that fall below the midpoint. These skills are: (1) Calculus, (2) Exponents/Logarithms, (3) Significant Figures, and (4) Scientific Notation, none of which are addressed in the QuaRCS. We note, however, that many science educators choose to emphasize these skills in their courses, scientific notation and significant figures in particular, despite their low degree of perceived importance.

Mathematics/numeracy educators consistently rank *all* skills as more important than science educators in *every* context, including the context of *science* literacy. Skills ranked by math educators at a full category, or more, higher than science educators for their importance to science literacy are: (1) Algebra, (2) Using Numbers in Writing, (3) Unit Conversions/Dimensional Analysis, (4) Making Graphs and (5) Statistics. Regarding the importance of quantitative skills in life, the differences in opinion were smaller on average, with only two standouts: (1) Making Graphs, and (2) Using Numbers in Writing. The fact that the differences in ranking are unique to certain contexts and vary according to the skill in question is interesting and worthy of followup.

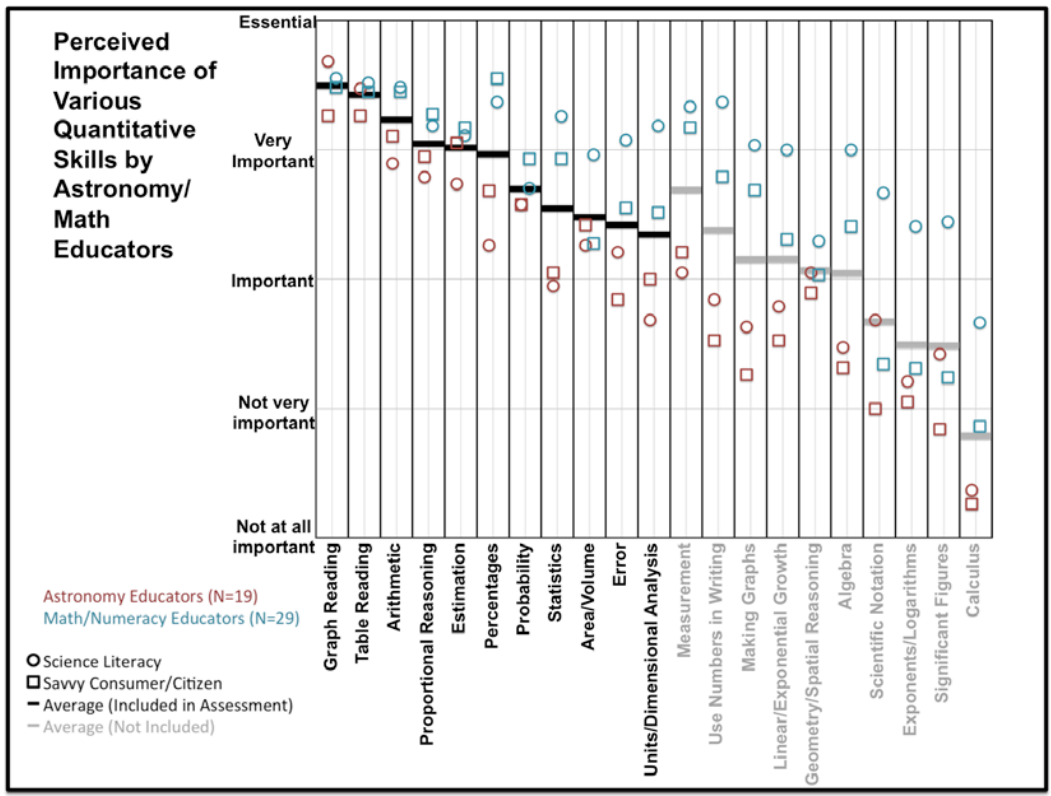

**Figure 2.** The perceived importance of various quantitative skills to science literacy (circles) and being a savvy consumer/citizen (squares) among Astronomy Educators (red symbols) and Math/Numeracy educators (blue symbols).

There are five skills in the list that stand out as being the most highly and consistently ranked by both math and science educators for science literacy and for educated citizenship. These are, in order of perceived importance: Graph Reading, Table Reading, Arithmetic, Proportional Reasoning, and Estimation. All five of these skills are assessed in the QuaRCS.

We have also chosen to draw from certain additional skills that are ranked in the "Very Important" region. The skills in this region, again in order of their perceived importance, are: Percentages, Measurement, Probability, Statistics, Area/Volume, Error, Using Numbers in Writing, and Dimensional Analysis/Unit Conversions. We include all of these skills in the QuaRCS except Measurement and Using Numbers in Writing. These are excluded because, though they are important and ranked highly, we felt that they would be more difficult to assess in a multiple-choice online assessment format.

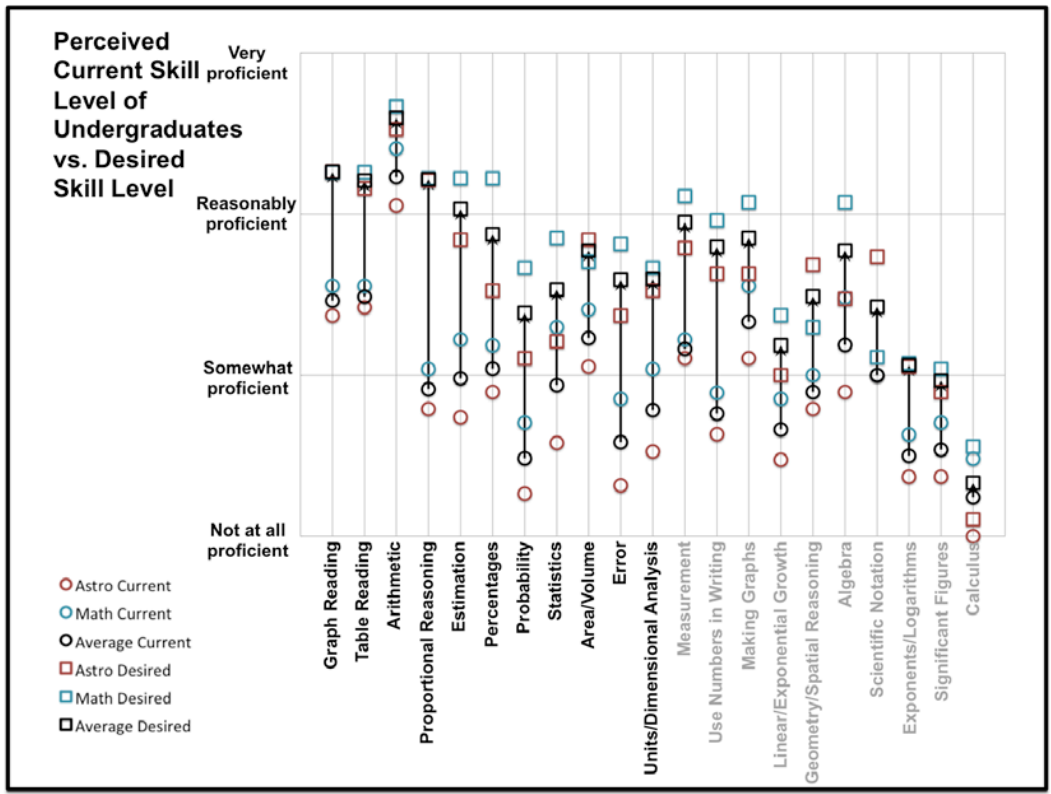

**Figure 3.** Desired (squares) and perceived current (circles) student proficiency levels for various quantitative skills among math (blue) and Astronomy (red) educators.

We also asked participating instructors to rate typical student ability levels at course entry, and the instructor's desired ability level for students. All but nine of the 48 participants specified that they were considering introductory courses for non-majors when answering these questions. The differences between these perceived and desired ability levels are shown in Figure 3. On the whole, instructors in both disciplines desire improved student skills in **all** areas, and they believe that correction of the disparity between perceived and desired proficiencies would have a significant effect on their course structure. Science educator responses to the followup question "How would your class be different if your students came in with this (desired) level of proficiency" included the following.

- We could explore not only deeper into topics, and a greater breadth of topics, but we could explore topics more relevant to their everyday and future lives, and ones that students find more interesting.

- They would have a better grasp of how science is done, including what information to trust and what evidence is good evidence.

- The 'I'm not good at math and science' hump likely wouldn't exist and they wouldn't shut down before they even started.

- More students would describe their nonmajor science class as 'useful.'

The skills that we assess in our instrument, in order of the amount of improvement that educators believe is needed (from greatest to least), are: Proportional Reasoning, Estimation, Error Analysis, Probability, Percentages, Unit Conversions/Dimensional Analysis, Graph Reading, Table Reading, Statistics, Area/Volume, and Arithmetic. Particularly notable among the skills that are not included in our instrument is the desired increase in proficiency in Using Numbers in Writing. Although we are unable to assess this skill with a multiple-choice assessment, members of Carleton College's Quantitative Inquiry, Reasoning and Knowledge (QuIRK) initiative have done much to advance this cause (e.g., Grawe and Rutz 2009, Grawe et al. 2010).

## Assessment Development

The QuaRCS was developed over the course of eight semesters between 2010 and 2015 and has been administered to over 4,000 students in more than 40 courses. Appendix A provides a description of instrument evolution semester by semester. Most courses were general education college science courses, although the assessment was also administered to several other populations for validation purposes, as described in the next section on item analysis. The QuaRCS contains an equal mixture of demographic/attitudinal questions ($N$=25) and "real world" quantitative questions ($N$=25). This paper focuses on classical test theory analysis of the quantitative questions on the assessment, as well as our efforts to establish the reliability and validity of the instrument as a whole. In Paper 2, we will describe results from initial administrations, including statistics about the demographics of the general education science student population.

Unless explicitly stated otherwise, all QuaRCS results reported are for the Spring 2015 pre-semester cohort of general education science students ($N$=1480). This administration represents both our largest student sample, and uses the final version of the QuaRCS instrument, refined over the course of eight semesters. All students in this sample were in one of 17 general education science courses at a large research University in the Southwestern U.S. and completed the QuaRCS between January 14 and February 12, 2015 (0-4 weeks into a 13-week semester). Enrollment in these courses ranged from 46 to 250 students (mean=121), and the QuaRCS was completed by 37% to 100% of students enrolled in each course (mean=72%). The courses were in a wide range of disciplines – 10 Astronomy,[7] 3 Biology, 1 Environmental Science, 1 Atmospheric Science, 1 Hydrology, and 1 Speech Language and Hearing Sciences.

All 45 instructors of courses fulfilling University-level general education Natural Sciences requirements in the Spring 2015 semester were invited to participate in the study. Six instructors declined, 16 did not respond after two inquiries, and the remaining 23 instructors initially responded positively. Seventeen of these instructors eventually elected to assign the assessment, either for participation credit or for extra credit. Participation rates vary according to which credit option was chosen (mean=79% participation for credit, 56% for extra credit). Per our IRB protocol, the QuaRCS is never administered in a high-stakes (graded) environment. Students simply receive credit, whichever variety their instructor elects, if they complete the instrument.

## *Question Wording*

Questions were developed over the course of eight semesters by a panel of six science educators and education researchers. The questions were designed to reflect situations that students might reasonably encounter in the course of their daily lives (e.g., bills, cooking, election polls, home repair), and they were worded as concisely as possible to reduce cognitive load. The final question set, together with item statistics, is discussed in detail in the next section on item analysis.

Figure 4 shows histograms of student responses to the questions: (a) "Overall, how difficult were the questions in this survey," and (b) "In your everyday life, how frequently do you encounter situations similar to the problems in this survey." A majority of students (52%) indicate that the instrument is moderately difficult, 30% find it easy or very easy, and 17% find it difficult or very difficult. Although few students perceive the questions as reflective of situations that they encounter in *daily* life (10%), the proportions who believe they encounter such situations *weekly* (23%) and *monthly* (31%) are higher than the proportions who believe they encounter them only *yearly* (13%) or *almost never* (23%).

---

[7] The reasons for the overrepresentation of Astronomy here are twofold. First, as the primary authors of this study are Astronomers, the response rate among Astronomy instructors was much higher. Furthermore, Astronomy represents nearly one-third of the Tier 1 and 2 Natural Science courses offered at the University.

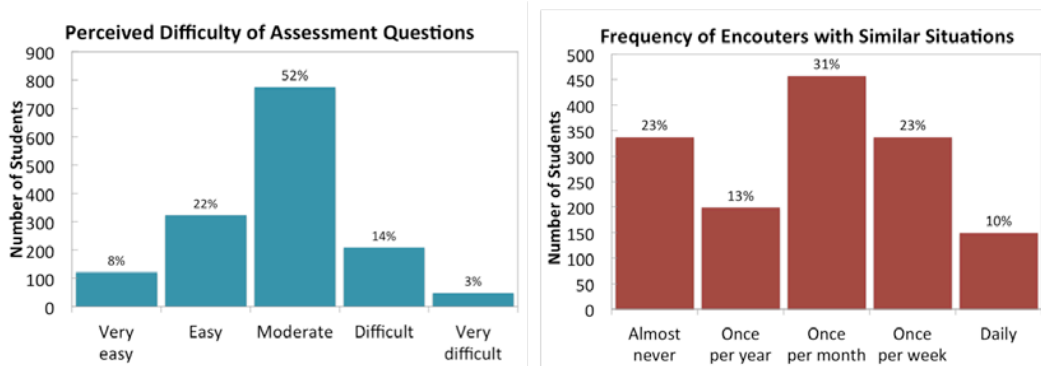

**Figure 4.** Histograms of student responses regarding the perceived difficulty of questions in the QuaRCS (left) and how frequently the students feel they encounter similar situations in their daily lives (right).

All quantitative questions were piloted in both multiple-choice and "free-response" forms in large general education Astronomy courses, and results of the free-response administrations are described in detail in the next section.

### *Development of Demographic and Attitudinal Questions*

The QuaRCS also asks students a series of low cognitive load demographic and attitudinal questions, which follow the quantitative questions in order to reduce the effects of stereotype threat. The purpose of these questions is to assess whether variables such as age, race/ethnicity, gender, major, and mathematical background are correlated with student performance on the assessment. Statistics regarding the development of demographic questions and correlations between these variables and performance are reported in Paper 2.

### *Inclusion of Confidence Rankings*

After each quantitative question on the assessment, students are asked to specify a degree of confidence in their answer. Figure 5 shows histograms of confidence rankings for the Spring 2015 cohort aggregated according to whether the quantitative question preceding it was answered correctly or incorrectly (1480 students, 25 questions each). Students answering correctly follow an expected distribution in confidence levels, with the majority either confident (29%) or very confident (51%) in their answers, and very few reporting having guessed (8%). Somewhat surprisingly, however, 44% of the respondents fall into the "confident" regime (15% very confident, 29% confident) when they answer ***incorrectly*** as well. This trend holds to within 10% for all earlier versions of the assessment and suggests that the QuaRCS is capable of measuring student awareness (or lack thereof) of their own quantitative abilities. In future administrations, we will study whether these distributions change over the course of a semester.

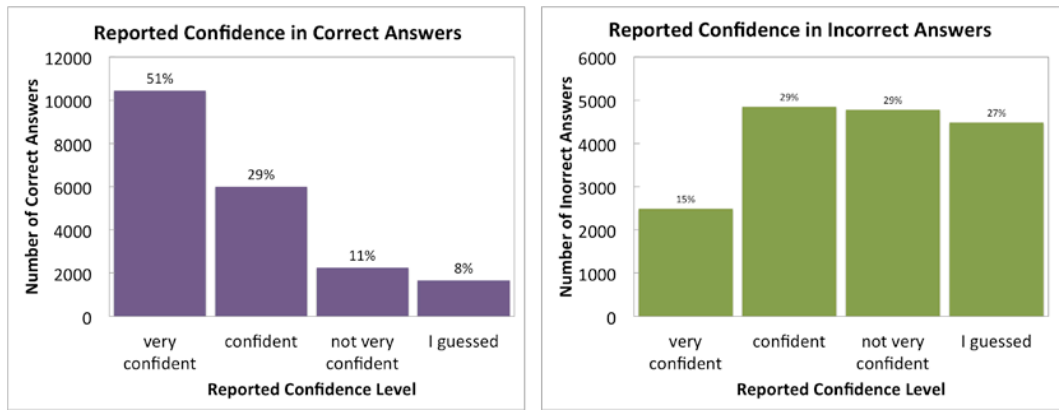

**Figure 5.** Histograms of student confidence in correct answers (left) and in incorrect answers (right). More than half of students answering questions incorrectly report that they are "confident" or "very confident" in their answers.

Given the import of these results and our desire to probe the question of whether student awareness of their own numerical abilities changes over the course of the semester, we have preserved these very low cognitive load followup questions to each quantitative question on the assessment for the time being. However, students often have negative affective reactions to this variety of repetitive followup question (Porter, 2004), so we may choose to remove them from future versions of the assessment.

## *Self-Reporting of Effort*

The low-stakes computer-based assessment format impacts student motivation, which can in turn affect students' scores (e.g., Wise and DeMars 2005, Sundre et al. 2008). As a result, "motivation filtering" (Sundre and Wise 2003) was utilized to mediate this effect   The last multiple-choice question on the QuaRCS appears as follows:

> *Knowing that this survey is being used for research to try to improve courses like yours and that your answer to this question will not be shared with your instructor, please honestly describe the amount of effort that you put into this survey.*
> *a) I just clicked through and chose randomly to get the participation credit*
> *b) I didn't try very hard*
> *c) I tried for a while and then got bored*
> *d) I tried pretty hard*
> *e) I tried my best on most of the questions*

We added this "effort question" to the assessment in Fall 2013 on the suspicion that early results were affected by increasing student apathy as the semester progressed. The distribution of student responses to the effort question on the Spring 2015 pre-semester assessment is shown in Figure 6.

Throughout the remainder of this paper, we will refer to the top two effort categories ("I tried pretty hard" and "I tried my best on most of the questions") as students who "devote effort" to the assessment. We will refer to students in the third effort category ("I tried for a while and then got bored") as students who "quit midway."[8]

We retain the data of students with low degrees of effort and those who quit midway for the majority of our analysis for several reasons. First, as most multiple-choice assessments do not include a similar question, the full unfiltered sample allows for direct comparison with other assessments. Secondly, correlations between self-reported effort and other assessment variables are very common. They are addressed in detail in Paper 2.

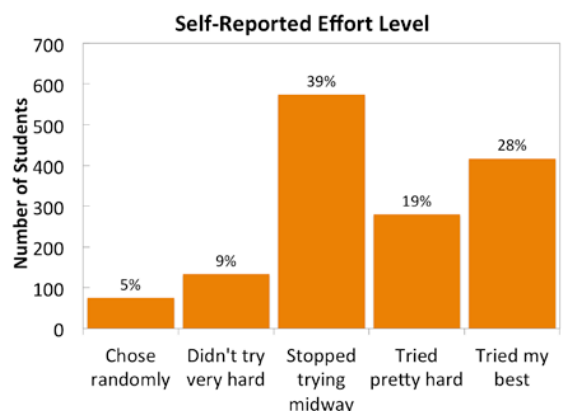

**Figure 6**. Distribution of student responses to the effort question. These responses (*N*=1480) are for the Spring 2015 cohort. Variations between administrations in students choosing each effort ranking have been <10%, despite significant modifications to the quantitative questions themselves.

## Assessment Format

After piloting the instrument as an in-class pen-and-paper assessment for two semesters (Fall 2010 and Spring 2011), we elected to move to an online format principally because the in-class time commitment proved a *severe* barrier to instructor recruitment. An online format allows instructors to assign the instrument to be taken outside of class, and it is therefore substantially easier to integrate into a course. Asynchronous online administration also allows students to complete the questions at their own pace. This flexibility is an important improvement over a pen-and-paper exam because the distribution in student responses times, shown in Figure 7, is exceptionally wide, from fewer than ten minutes to over an hour.

---

[8] These students did, in fact, complete the assessment in the sense that they selected an answer for every question. However, as discussed in detail in the section on assessment length, their performance on the first and second halves of the assessment differ in statistically significant ways.

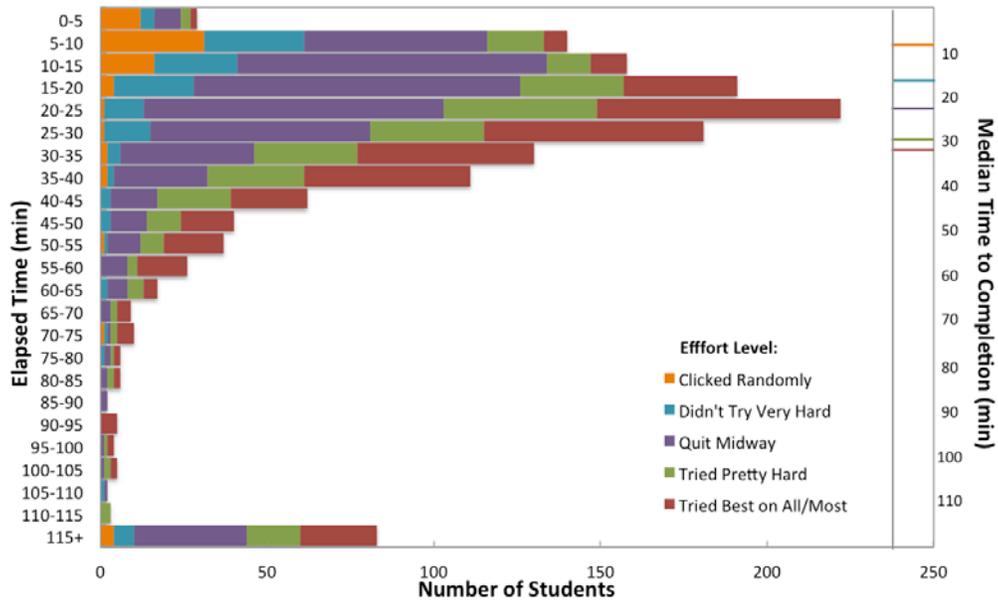

**Figure 7.** Distribution of time elapsed from start to completion of the Spring 2015 QuaRCS for the 1480 General Education science students. These data are binned in five-minute increments and according to self-reported effort level. The median time to completion for each effort level is indicated at right.

Online administration is not without disadvantages, however. The long tail of the time distribution shown in Figure 7, and in particular the cluster of completion times greater than two hours, suggests that some students are not actively engaging with the instrument throughout. They may be taking it in more than one sitting, multitasking, or idling. As the online data collection interfaces used[9] are unable to measure ***active*** engagement in that they record only the start time and completion time for each survey response, we cannot precisely quantify these effects. However, just 10% (*N*=152) of students took longer than 60 minutes to complete the Spring 2015 QuaRCS, and the mean score of this group (14.5, 58%) is less than 1 point higher than that of the entire sample (13.8, 55%). Some computer-based testing platforms are capable of measuring response times for individual items, and we will consider new platforms in the future. In particular, we are eager to investigate whether self-reporting of effort and score correlate with time on task.

---

[9] We used DatStat Illume until Fall 2014 and Qualtrics in Spring 2015.

The median times to completion are strongly correlated with effort level. Students who report having chosen answers randomly in order to obtain credit have a median completion time of 8.1 minutes, while students in the top two effort categories have median completion times of 29.4 and 31.3 minutes, respectively. With the reasonable assumption that students taking longer than two hours to complete the assessment have simply left it open on their computer, and that the actual distribution of time spent engaging actively with the assessment falls off sharply after 60 minutes, this range is well within that of a typical homework assignment in a college science course.

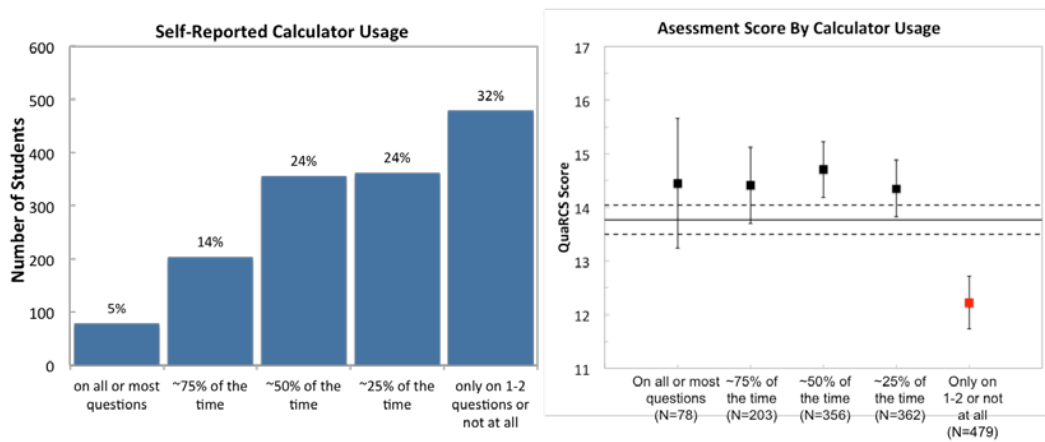

**Figure 8.** Left: Histogram of student responses from the Spring 2015 administration reflecting the self-reported usage of calculators. More than half of students report that they use a calculator on 25% or fewer of the questions. Right: Mean assessment score according to self-reported calculator usage. The solid horizontal line indicates the mean of the entire general education student population, and the dashed lines indicate 95% confidence intervals on that mean. Square symbols indicate the mean score for each population, and error bars are the 95% confidence intervals on those means. Only one group, those who use calculators the least (shown in red), deviates from the other groups in a statistically significant way, scoring ~2.5 points (10%) lower on the assessment than the other groups.

The online format also does not allow us to control calculator usage. We elected to include the following statement in the introduction to the online assessment: "These questions were designed to be answerable without a calculator, but you are welcome to use one if you choose." We ask students to self-report their usage, and this distribution is shown in Figure 8. Although the responses vary significantly, a majority eschew calculators on 75% or more of the questions (32% using one on just 1-2 questions or not at all, and 24% using a calculator 25% of the time or less). The group that reports using calculators the least is also the lowest-performing group, as shown in Figure 3b, deviating from the rest of the population by more than 2 points ($p<0.001$). However, calculator usage varies across effort groups at $p<0.001$, with the students trying the least hard also using calculators the least.

Given the relatively low proportions of (a) students clearly dividing their attention between the instrument and other activities, and (b) overreliance on calculators, we believe that the benefits of the online format outweigh the problems. As further validation of this choice, we note that the average score on the QuaRCS dropped by just ~3% between the Spring 2011 paper-format assessment and the Fall 2011 online version, although the majority of the questions were substantively unchanged.

## *Content Alignment*

Measuring 11 quantitative skills in a single assessment is an ambitious endeavor, and test fatigue (PISA 2000, Wise et al. 1989) was a concern. We took advantage of overlap between skill categories, and each of our questions assesses multiple skills simultaneously.

In order to validate assignment of specific quantitative skills to QuaRCS questions, we had 15 numeracy experts (a mixture of numeracy and mathematics educators) classify questions according to which skill(s) they assess. Multiple skill selections for each quantitative question were encouraged, and a brief description of each skill was provided in order to ensure that skill categories were interpreted equivalently by all experts. A skill was assigned to a question if seven or more experts classified it as such. Due to a dearth of questions classified as assessing "probability" in this analysis, we combined this category with statistics for a final set of ten numerical skills.

Table 2 gives the final ten skill categories ordered according to perceived importance by math and science educators, the description that was provided to experts for each skill, the number of QuaRCS questions classified as falling into these categories, and the range of difficulties. Here and elsewhere in this paper, "difficulty" is the $p_D$-value[10] of the question, or proportion of students answering correctly.[11]

All but three skills span wide ranges in difficulty. We made several attempts to write "easy" Statistics/Probability and Error questions and a "hard" Area/ Volume question, but were unsuccessful. The narrow range of $p_D$-values for these skills may be a reflection of an underlying level of comfort or familiarity (e.g., comfort with Area/Volume, and discomfort with Statistics/Probability and Error), or we may merely have failed to hit upon the right context. Further work is needed to determine the root cause of the narrow range in difficulties for these skills.

---

[10] This is generally referred to as simply the "p" value for a question in Classical Test Theory, but we've given it the subscript D in this case to distinguish it from the $p$-value denoting statistical significance.

[11] e.g., $p_D$=0.1 means 10% of students answered correctly. A question with a high $p_D$-value means that students score high on it. Difficult questions have low $p_D$-values.

**Table 2**
**QuaRCS Skill Categories**

| Skill | Definition | Abbrev. | $N_{Quest}$ | $p_D$-value range |
|---|---|---|---|---|
| **Graph Reading** | Read, interpret or extrapolate graphical data. | GR | 5 | 0.22-0.76 |
| **Table Reading** | Read and interpret information presented in tabular form. | TR | 3 | 0.35-0.80 |
| **Arithmetic** | Add, subtract, multiply or divide two or more numbers. | AR | 21 | 0.24-0.80 |
| **Proportional Reasoning** | Compare two or more numbers, rates, ratios, fractions. | PR | 13 | 0.24-0.75 |
| **Estimation** | Approximate an answer or choose the closest value to a precise calculation. | ES | 4 | 0.24-0.76 |
| **Percentages** | Compute or compare percentages | PC | 5 | 0.28-0.73 |
| **Statistics and Probability** | Statistics= interpretation of data, including distributions and descriptive statistics (mean, median, mode, etc.). | SP | 6 | 0.22-0.59 |
| | Probability = compute odds or risk or determine the most likely outcome. | | | |
| **Area and Volume** | Compute or compare areas or volumes | AV | 5 | 0.48-0.68 |
| **Error** | Evaluate uncertainty in graphs or numbers | ER | 4 | 0.22-0.36 |
| **Unit Conversions and Dimensional Analysis** | Unit Conversions = Use the relationship between two or more units to transform one number into another. | UD | 6 | 0.30-0.75 |
| | Dimensional Analysis = Draw inferences about the relationship between two or more quantities based on the units attached to them. | | | |

The final question stems, their categorization, and various statistics about difficulty, discrimination and reliability are given below under Item Analysis.

## *Assessment Length*

To study the effect of assessment length and fatigue on student responses, we administered a reverse-ordered version to half of the students in one large lecture class in Fall 2013. The pre- and post-semester instruments were completed by 91, and 90 students, respectively, and approximately half of these students were assigned the reverse-ordered instrument in each case ($N$=43, 42).

As a first measure of the effects of test fatigue, we completed an analysis of variance test on scores for two large[12] blocks of questions (11 questions each). We found that students who devoted effort scored equally well on a given block whether it appeared in the first or second half of the assessment. This leads us to conclude that fatigue does not figure prominently into item difficulties ***among students who expend effort on the assessment.***

---

[12] In order to decrease the effects of small sample sizes.

We note, however, that we have excluded all students who reported a lower degree of effort than "I tried pretty hard" ($N_{pre}$=30, $N_{post}$=42). Particularly important to the question of instrument length are students who report quitting midway ($N_{pre}$=29, $N_{post}$=27), for whom there are often, though not invariably, statistically significant ($p<0.05$) differences in score for a given question or block of questions according to whether it was encountered early or late in the assessment. To prevent this effect from altering the item statistics for certain questions disproportionately, we randomized the order of all but the first three questions in the assessment for the collection of final instrument statistics. We note, however, that all questions are still subject to the effects of student apathy and that randomization merely spreads the effect roughly equally between questions.

Also relevant in determining the appropriate instrument length is the relative *number* of students who report quitting midway. This population has consistently composed 25–40% of students on all versions of the assessment. Because the effort question was devised after moving to online administration, it is not clear to what extent this large proportion is a result of the self-timed format. We will recruit several instructors to administer a pen-and-paper assessment in future semesters in order to probe this question further.

The large proportion of students who quit midway is far from ideal; however, we chose to maintain the long form of the instrument in order to assess a broader range of skills reliably and in multiple contexts. In future work, we will pilot a shortened version of the QuaRCS drawing only from the most highly ranked skills from the educator survey (described in the previous section) in an effort to reduce this population. A shorter assessment is likely to be less reliable and will certainly have narrower content coverage, but students may be better able to maintain effort throughout. However, given the positive correlation we have found between attitudes toward mathematics and effort level (Paper 2), it is not clear whether a shorter assessment will result in a lower proportion of students in this category. It may well be that the fraction of the population with poor attitudes toward mathematics will always show a low level of persistence when it comes to numerical questions.

## *Matched Data*

Both the Spring 2012 and Fall 2013 QuaRCS assessments included pre- and post-semester administrations. Pre/post data will be the norm in future semesters and will be essential for studying the effectiveness of various interventions during the broader QuaRCS study (supported in part by an NSF Transforming Under-graduate Education in STEM grant, and described in Follette and McCarthy 2012). In order to match data, we ask students to provide names at each administration, and we used these data to match pre- and post-semester results.

Per our IRB protocol, names and corresponding scores are kept confidential and instructors are provided only lists of students who completed the instrument, not individual results.

# Item Analysis

## *Free-Response Items*

All quantitative questions on the QuaRCS have been vetted in "free-response" form as open-ended versions of the multiple-choice question stems. The questions were distributed in large general education Astronomy courses, and students were given approximately 5–10 minutes at the beginning of a class period to complete them for participation credit. Administration took place over the course of three semesters (Spring 2012, Fall 2013, and Fall 2014) as questions were drafted and revised. Responses were hand coded according to a rubric that assessed (a) whether the student arrived at a correct answer, (b) whether work was shown, (c) whether that work was correct, and (d) what type of work was done (written explanation, long division, drew diagram, etc).

This analysis served several purposes for question development and instrument validation. First, asking the questions in an open-ended format allowed us to generate compelling multiple-choice distractors based on common mathematical misconceptions and mistakes. The majority of QuaRCS questions had at least one distractor added through this analysis.

A second benefit of the free-response analysis was that it illuminated instances where words or phrasing were unclear, imprecise, or could be misinterpreted. Any substantial wording changes, as well as minor wording changes that were misinterpreted by more than 5% of respondents, were given again in free-response form to ensure that the problem was corrected.

A final benefit of the free-response testing was to ensure that students were reaching the correct answer through legitimate quantitative reasoning and not stumbling upon it through error. For several questions, open-ended responses revealed that students could arrive at the correct answer serendipitously, despite an incorrect quantitative approach. In all cases where more than one student in the sample arrived at the correct answer through incorrect reasoning, the question was adjusted and rerun in free-response form in order to ensure that the problem was corrected.

In all, 35 questions were developed and tested in free-response form. Of these, ten were rejected because of inability to clarify the wording sufficiently, topical redundancy, or low discrimination, and seven were revised and retested before incorporation into the final instrument. The final wordings of the multiple-choice question stems are given in Table 3.

**Table 3. QuaRCS Items and Item Statistics**

Items and item statistics for the final version of the QuaRCS assessment. Question numbers are arbitrary, as the order of question blocks are randomized on the assessment, although they match the question numbering of Figure 9. Blocks of questions are separated by triple lines. The skill categorization for each question based on expert analysis is given in Column 3, where abbreviations match those of Table 2. Multiple-choice item difficulties and discriminations are given for both the General Education ("Gen Ed") science student population (Columns 5-6) and for a Differential Equations ("Diff EQs") STEM major course (Columns 7-8). Questions falling within the commonly accepted ranges for difficulty ($0.2 < p_D < 0.8$) and discrimination ($>0.3$) are bolded in the table. Questions not fulfilling these criteria are described in detail in the text. For comparison, item difficulties for the same questions given in free-response form to the general education science student population are given in Column 4. In several cases, the final question was also administered in multiple-choice form with instructions for students to explain their answer choice (indicated with [+] symbol).

| No | Question Text | Skills | Gen Ed Free Response Difficulty | Gen Ed Multiple Choice Difficulty (N=1480) | Gen Ed Discrimination | Diff. EQs Difficulty (N=261) | Diff. EQs Discrimination |
|---|---|---|---|---|---|---|---|
| 1 | You have a rectangular fish tank that's 10 inches tall, 20 inches wide, and 15 inches deep. If the volume of one gallon of water is 231 cubic inches, then how many gallons are required to fill the tank? | AR AV UD | 0.70 (N=56) | **0.68** | **0.32** | 0.91 | **0.46** |
| 2 | Your grocery store has a 20 ounce jar of peanut butter for $4.00, and a 45 ounce jar for $9.00. Which purchase will get you the best price per ounce? | PR AR | 0.93 (N=61) | **0.72** | **0.42** | 0.94 | **0.45** |
| 3 | A college that typically has 50,000 students experiences an increase in enrollment to 55,000 students. By what percentage did enrollment increase? | AR PC | 0.77 (N=93) | **0.53** | **0.43** | 0.86 | 0.28 |
| 4 | According to the graph, what was the approximate population of City X in 1980? | GR ES | 0.91 (N=59) | **0.76** | **0.33** | 0.91 | **0.32** |
| 5 | If the current population growth rate continues, which is the best estimate for the population of City X in the year 2050? | GR ES PR | 0.40 (N=102) 0.71[+] (N=48) | **0.43** | 0.20 | **0.62** | **0.33** |
| 6 | Based on this graph, compare the population growth rates (i.e. increase in number of people per year) before and after 1970. | GR PR AR ES | 0.36 (N=88) | **0.24** | 0.27 | **0.54** | **0.34** |
| 7 | Imagine you have already filled a measuring cup (like the one shown above) with the amount of peanut butter in the recipe and you want to add the correct amount of shortening **on top of** it. Which line on the measuring cup should you fill to with shortening? | AR | 0.85 (N=52) | **0.70** | **0.43** | 0.99 | **0.32** |
| 8 | If your measuring cup has ounces on the side instead of cups, which line should you fill to when measuring the flour? There are 8 ounces in 1 cup. | UD AR | 0.87 (N=35) | **0.75** | **0.46** | 0.92 | **0.41** |
| 9 | You have only a **½ Tablespoon** measuring spoon. How much should you fill it to get the correct amount of baking soda? There are 3 teaspoons in 1 tablespoon. | AR UD PR | 0.41 (N=33) 0.67[+] (N=39) | **0.30** | **0.36** | 0.62 | **0.45** |

| No | Question Text | Skills | Gen Ed Free Response Difficulty | Gen Ed Multiple Choice Difficulty (*N*=1480) | Gen Ed Discrimi-nation | Diff. EQs Difficulty (*N*=261) | Diff. EQs Discrimi-nation |
|---|---|---|---|---|---|---|---|
| 10 | How many total injuries (including deaths) were sustained at Resort Y during this time period? | TR AR | 0.89 (*N*=55) | **0.80** | **0.34** | 0.92 | **0.38** |
| 11 | What were the chances of a randomly-selected skier sustaining an injury of any kind (minor, severe or death) while at Resort Y during this time period? | TR SP AR PC | 0.76 (*N*=50) | **0.58** | **0.39** | 0.85 | **0.41** |
| 12 | What proportion of **severely injured** skiers at Resort Y during this time period were **intermediate** skiers? | TR AR PR SP | 0.70 (*N*=78) | **0.35** | **0.42** | 0.63 | **0.33** |
| 13 | The graph above shows the **predicted** viewership of three television shows in two cities based on a poll of a small number of residents in each city. The poll has a reported error of 25%, shown as vertical error bars. Which of the following statements about the predicted viewers of Show A is most accurate? | GR ER SP | 0.26 (*N*=50) 0.56[+] (*N*=50) | **0.36** | 0.22 | 0.61 | **0.27** |
| 14 | Which of the following predictions can be made based on the information (including errors) shown in the graph? Prediction 1: In City 2, more people will watch Show B than Show C Prediction 2: In City 1, Show C will have the smallest viewership Prediction 3: None of the three shows (A, B or C) will be equally popular in Cities 1 and 2 | GR ER SP | 0.21 (*N*=42) 0.24[+] (*N*=50) | **0.22** | **0.36** | 0.53 | **0.34** |
| 15 | You purchased 100 square feet of solar panels for your roof. However, your local Homeowner's Association requires that solar panels not be visible from the road. You decide to put solar panels on the roof of your shed instead. The shed has a flat 5 foot by 5 foot roof. Complete the following sentence: "To produce the same amount of power as your original design, you need to buy panels that produce _______ times more power per unit area than your original panels." | AR PR AV | 0.82 (*N*=56) | **0.62** | **0.47** | 0.89 | **0.43** |
| 16 | If you cover the shed with your original panels, how many **more** of the same size sheds would you have to put up in your backyard in order to fit the rest of the panels? | AR PR AV | 0.83 (*N*=52) | **0.48** | **0.50** | 0.84 | **0.57** |
| 17 | Your cable bill is $36 per month from January 1 through September 30 and then doubles to $72 per month starting October 1. What is your average monthly bill over the course of the entire calendar year (January-December)? | AR PR SP | 0.78 (*N*=68) | **0.59** | **0.37** | 0.80 | **0.35** |
| 18 | If you place $10 under your mattress every day for the next 40 years, approximately how much money will you have? | AR | 0.89 (*N*=37) | **0.74** | **0.30** | 0.90 | 0.20 |

| No | Question Text | Skills | Gen Ed Free Response Difficulty | Gen Ed Multiple Choice Difficulty ($N$=1480) | Gen Ed Discrimination | Diff. EQs Difficulty ($N$=261) | Diff. EQs Discrimination |
|----|---------------|--------|---------------------------------|----------------------------------------------|----------------------|--------------------------------|---------------------------|
| 19 | A newspaper conducts a survey and predicts that in the local election between Candidates A and B, Candidate A will receive 60% of the votes. The newspaper estimates the error in this prediction to be 5%. If the newspaper conducts another survey with 400 participants, how many people can report that they will vote for Candidate A for the result to be consistent with the original conclusion (that Candidate A will receive 60% of the votes with 5% error)? | ER AR PC SP | 0.64 ($N$=72) | **0.28** | **0.48** | **0.61** | **0.42** |
| 20 | Several days later, the newspaper conducts another survey with 300 new participants. What is the **minimum** number of votes that Candidate A can receive in this new survey in order to be consistent with the original prediction (that Candidate A will receive 60% of the votes with 5% error)? | ER AR PC SP | 0.80 ($N$=82) | **0.36** | **0.46** | **0.71** | **0.44** |
| 21 | You want to carpet a 15 foot by 20 foot room. You have two carpet options to choose from. One is $1.50 per square foot and the other is $3.00 per square foot. How much more will your total bill be if you choose the more expensive carpet rather than the cheaper one? | AV AR PR | 0.91 ($N$=102) | **0.62** | **0.49** | 0.90 | **0.39** |
| 22 | To carpet your 15 foot by 20 foot room and a hallway that is 4 feet by 12 feet, how much total carpet do you need? | AV AR | 0.95 ($N$=88) | **0.66** | **0.54** | 0.94 | **0.55** |
| 23 | If one scoop of lemonade powder is needed for every 12 ounces of water, then how many scoops should you add to **3 gallons** of water to make it into lemonade? 16 ounces = 1 Pint 2 Pints = 1 Quart 4 Quarts = 1 Gallon | UD AR PR | 0.70 ($N$=83) | **0.50** | 0.28 | 0.76 | 0.29 |
| 24 | A sweater that was originally $100 is on sale for 30% off. Which of the following coupons should you use to get the lowest final price? | AR PC PR | 0.79 ($N$=52) | **0.73** | **0.35** | 0.89 | 0.21 |
| 25 | You drove 200 miles on 11 gallons of gas. Which of these is closest to the number of miles per gallon that you got? | AR ES PR UD | 0.95 ($N$=93) | **0.75** | **0.48** | 0.95 | **0.46** |

## Administration to Experts

Another key aspect of QuaRCS development was establishing its capability to distinguish varying levels of quantitative literacy. To assess this potential, we administered the instrument to groups of "experts" (science and numeracy educators) at two different points during development.

**Fall 2014 Expert Administration**. In the summer of 2014, the most recent version of the assessment (Fall 2013) was administered to a group of 34 science ($N$=17) and numeracy ($N$=17) educators, who completed it as though they were students. The faculty score distribution (M=22.8, SD=1.9) is markedly different from that of even the highest-scoring group of students (those who devote effort, $N$=251, M=17.1, SD=4.1). Using an independent samples *t*-test, we find that these populations are different at $p$<0.001.

These results provided early evidence that the instrument is capable of distinguishing experts (the "quantitatively literate") from novices, but it also helped us to identify three problematic questions. When comparing the difficulty of individual questions for instructors and students, three questions stood out as being anomalous. Two were answered correctly by fewer than half of the experts, and a third was answered correctly by students more often than by experts, though both groups answered correctly more than 80% of the time.

Through interviews, it became clear that experts were often misinterpreting the simplified wording of the first two questions, in both cases complicating the problem more than was intended. We removed these items from the assessment, as advanced students are likely to make similar misinterpretations. In the case of the third question (number 25 in Table 3), we determined that the answer choices were too closely spaced for a non-calculator user to distinguish quickly between the correct and incorrect response. It would seem that experts were quick to estimate the answer as the problem intended, although approximately one in five estimated wrong. As revealed through free-response administration, students were more likely to actually "do the math," whether through longhand arithmetic or the use of a calculator and, therefore, answered correctly more often. To correct this minor problem, we spaced the distractors farther from the correct answer to more strongly encourage students to estimate, as well as to discourage experts from estimating incorrectly.

If the three problematic items are removed, then the average score on this assessment among experts becomes 20.9/22 (95%) and the standard deviation falls to 1.3 (6%). Among students who devoted effort, the average on the same 22 questions is 15.3 (70%), and the standard deviation is 4.2 (19%).

**Spring 2015 Expert Administration.** The results of administering the instrument to the first round of experts, in tandem with analysis of free-response questions and classical test theory statistics from early versions of the instrument, led to modifications before final implementation. Ultimately, 17 of the questions from the Fall 2013 instrument remained substantively unchanged into the final instrument. However, eight questions were replaced due to low discrimination, topical redundancy, or both. All new questions were vetted in both free-response and multiple-choice formats in the Fall of 2014 before the final administration in Spring 2015.

We administered the final version of the instrument to an additional 12 experts in January 2015, and item difficulties for the final question set are shown in Figure 9. This second group of experts achieved a mean score of 23.4 (94%) with a standard deviation of 1.2 (5%) on the QuaRCS. This result is also statistically significantly different from the scores of general education science students at the $p<0.001$ level.

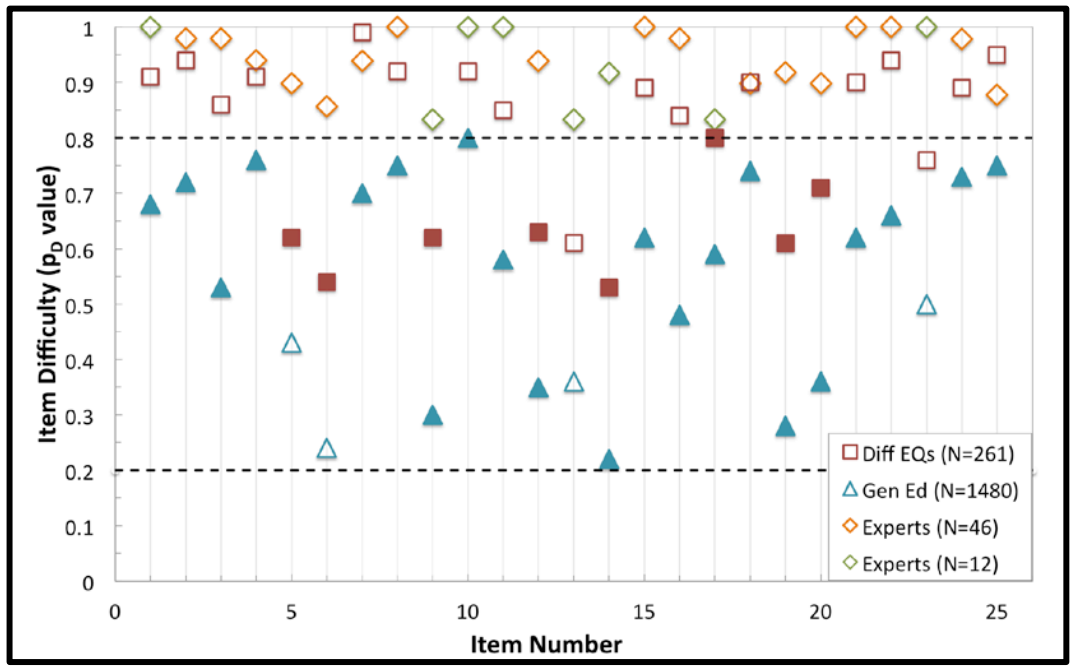

**Figure 9.** Difficulty ($p_D$) values for individual QuaRCS items for general education science students (blue triangles), Differential Equations students (red squares) and experts (orange and green diamonds). QuaRCS questions were administered to groups of experts twice ($N$=36, 12). In cases where the question remained substantively unchanged between expert administrations, the results have been aggregated (orange diamonds). New or significantly altered questions, administered only to the second group of experts, are shown with green diamonds. In all cases, filled symbols indicate items with discrimination values greater than 0.3, which is the commonly accepted cutoff for high quality multiple choice items. The difficulty range from 0.2 to 0.8 is considered ideal for assessments, and these cutoffs are shown with horizontal dashed lines. All of the questions lie in this range for the general education student population. Items with low discrimination values are discussed in detail in the text.

## Item Statistics

Throughout development, we evaluated the quality of the instrument using Classical Test Theory (CTT), for which it is assumed that a student's score on an assessment reflects their (unmeasurable) true score plus a random error (Allen and Yen 2002). The evolution of instrument statistics such as mean, standard deviation, and Cronbach's $\alpha$[13] during development are reported in Table 4. Appendix A describes changes between instrument versions in detail.

**Table 4**
**Instrument Evolution**

| Form | N Students | N Questions | Mean (%) | Standard Deviation (%) | Cronbach's $\alpha$ |
|------|-----------|-------------|----------|------------------------|----------------------|
| Spring 2011 | 68 | 10 | 66 | 22 | 0.692[14] |
| Fall 2011 | 190 | 22 | 63 | 19 | 0.767 |
| Spring 2012 | 574 | 22 | 58 | 18 | 0.749 |
| Fall 2013 | 518 | 25 | 59 | 19 | 0.801 |
| **Fall 2013\*** | **343** | **25** | **63** | **20** | **0.825** |
| Fall 2014 | 276 | 25 | 55 | 23 | 0.865 |
| **Fall 2014\*** | **166** | **25** | **62** | **24** | **0.885** |
| Spring 2015 | 1480 | 25 | 55 | 21 | 0.843 |
| **Spring 2015\*** | **906** | **25** | **60** | **22** | **0.868** |

Evolution of instrument statistics from Spring 2011 through Spring 2015 as the QuaRCS was developed and refined. Reported statistics for each version of the assessment include the number of students (*N* Students), the number of questions (*N* Questions), the mean score, the standard deviation in score, and the Cronbach's $\alpha$ value. The effort question described in the text was implemented in Fall of 2013, and from this point forward we also give instrument statistics with the population of students who "quit midway" through the assessment excluded. These rows appear in bold and are marked with asterisks. Arguments regarding the utility of separating this group are given in the text, under Item Statistics.

For all administrations, the mean score on the QuaRCS among general education science students was in the 55–65% range and the standard deviation was very high (18–24%), indicating a low overall ability and a wide range of abilities for this population. Cronbach's $\alpha$ has remained at or above 0.7, considered in the acceptable range for a multiple-choice assessment (George 2003), since initial administration, and reaches 0.87 for the final version of the QuaRCS.

Values in Table 4 are reported for both (a) the complete unfiltered datasets and (b) with students who "quit midway" filtered out. The statistics for the full sample allow for a more direct comparison with other multiple-choice assessments, while the filtered subsample highlights the fact that including students whose effort wanes midway through the assessment results in an overall decrease in Cronbach's $\alpha$. This correlation does not occur when excluding ***any*** of the other effort categories, as other students are consistent in their responses throughout.

Statistics for individual items are reported in Table 3. These statistics include item difficulties for individual questions in both free-response and multiple-choice formats. The $p_D$-values for free-response questions are an average of 0.31 higher than the same multiple-choice question. This difference is perhaps to be expected, as free-response questions were administered one at a time during class, so motivation was likely high and fatigue negligible. Lack of multiple-choice options in this format also makes guessing more difficult. In general, difficulty values for multiple-choice assessments are considered acceptable when they lie between 0.2 and 0.8 (Bardar et al. 2006, Schlingman et al. 2012), and all of our questions lie within this range for the General Education science student population.

---

[13] Cronbach's $\alpha$ is a CTT measure of the internal consistency of an assessment. It is a number between 0 and 1, where high values indicate high reliability

[14] this value was computed only for the four questions that were common to all versions of the assessment given this semester

Discrimination values for each question are also given in the table, and these values describe how well an item distinguishes between high-scoring and low-scoring individuals. [15] Conventionally accepted values for this parameter lie between 0.3 and 0.7 (Bardar et al. 2006). The majority of items on our assessment lie in this range; however, three of the five graph-reading questions (items 5, 6 and 13) have values in the 0.2–0.3 range. It is not entirely clear why these low results occur. Perhaps graph reading is different enough from the other skills assessed on the QuaRCS that student scores in this area do not track with their overall score, or perhaps these questions are problematic in some way that our initial analysis did not identify. We will investigate this finding further in future work. We note that all three low-discrimination items are also difficult ($p_D$ values of 0.43, 0.24, and 0.36), and lower discrimination values are not uncommon for high-difficulty items (Schlingman et al. 2012).

Item 23 also shows a low discrimination (0.28) in the Spring 2015 data; however, we believe this result is artificial because we neglected to bold and underline "three gallons" when the final version of the assessment was transferred to a new administration platform. The same question appeared in the Fall 2014 assessment with a discrimination of 0.46, and the only difference was the bolding of the phrase "three gallons." This error will be corrected in future administrations.

## Population validation

We administered the instrument to several additional student populations in order to assess its ability to measure quantitative literacy in populations besides general education science courses at large research universities. For the final version of the instrument, these administrations were for two general education science courses at a minority-serving Community College in the Southwest ($N$=48) and a large mid-level math course (Ordinary Differential Equations, $N$=261) at the same large research University in the Southwest as the general education cohort described in the bulk of this paper. We also administered an earlier (Fall 2013) version of the instrument to a second-semester Algebra-based physics course at a small liberal arts institution in the Midwest ($N=80$) and to a general education Astronomy course at the same Community College in the Southwest ($N=20$).

---

[15] Discrimination is defined as the difference in $p_D$ values between high- and low-scoring individuals. Generally, these are defined as the top and bottom 27% of the population, respectively.

**Table 5**
**Performance of Various Student Populations on the QuaRCS**

| Course | Institution | Instrument Version | $N_{students}$ | Mean | Std Dev | C α |
|---|---|---|---|---|---|---|
| Second Semester Algebra-Based Physics | Midwest Liberal Arts College | Fall 2013 | 80 | 83% | 8% | 0.402 |
| General Education Astronomy | Southwest Community College | Fall 2013 | 20 | 60% | 17% | 0.755 |
| Differential Equations | Southwest Research University | Spring 2015 | 261 | 80% | 16% | 0.822 |
| General Education Astronomy | Southwest Community College | Spring 2015 | 48 | 63% | 16% | 0.766 |

Table 5 lists statistics for each of these additional populations, including Cronbach's α. Generally speaking, lower Cronbach's α values are to be expected for all of the test populations. Small-number statistics will affect the Community College course statistics, and low item difficulties among the predominantly STEM major population of the other two courses should also decrease reliability. Indeed, the value of the statistic is lower in all cases than for the large University general education cohort, but it is still well within the acceptable range for quality assessments in all but one case.

The second-semester algebra-based physics course falls outside of the range for quality assessments, with a Cronbach's α value of just 0.40 (though it reaches 0.55 through the exclusion of the five questions with the lowest discrimination). It is not clear why Cronbach's α is so much (0.4) lower for this course than for the Differential Equations course, and more work is needed to understand this population, including analysis of demographic, attitudinal and effort data. The assessment may be valid only for STEM major courses in certain situations; this may be an effect of a small number of outliers in this relatively small sample; or transcription errors may have occurred when the instructor moved our assessment to a different online platform. We do not have access to the raw data for this course, only scored results, but we will explore the validity of the QuaRCS among STEM major courses further in future work.

The high level of reliability even in very small Community College General Education science courses is consistent with our assertion that the QuaRCS is a reliable assessment for General Education science students regardless of institution, and we will continue to test this assertion as the QuaRCS is adopted more widely. This is not to say, however, that the General Education science student population at the minority-serving Community College is identical to the University General Education population that we used for our validation study.

Interestingly, the Community College population deviated in score **and** in self-reported effort level from both the Tier 1 and Tier 2[16] University populations. The Community College students scored, on average, 2.3 and 1.5 points higher than Tier 1 and Tier 2 University students, respectively, and these differences in score were statistically significant in both cases ($p$=0.002 and 0.051). This result appears to be due to the Community College population's trying harder on average. Scores among "high effort" students are statistically indistinguishable between the two institutions, but the overall effort response distributions are significantly different from one another ($p$<0.001). **None** of the 48 Community College students fell into the "low effort" category. Although the sample is too small at this point to draw robust conclusions from this result, it is intriguing and will be investigated further in future work.

Given the large sample size, we also paid particular attention to the results of the Spring 2015 Differential Equations course. The 261 students in this course were overwhelmingly STEM majors (93% Engineering, Computer Science or Mathematics, 8% Science, 3% Health Professions and every other major <2%[17]), and only 13% of them were freshmen. This group scored an average of 20.0 (80%), with a standard deviation of 4.1 (16%), and 92% of them fell into the top two effort categories, marking them as a significantly different population than the general education science students ($p$<0.001).

Difficulty and discrimination measures for individual items are provided in Table 4 for the Differential Equations student population. On average, $p_D$-values for QuaRCS items are 0.25 higher (students answer correctly 25 percentage points more often) for this population than among general education science students at the same University, but only 0.01 lower in discrimination. Eight QuaRCS questions, highlighted with filled red squares in Figure 9, were within the commonly accepted range of difficulty **and** discrimination for quality assessments, and many additional low-difficulty ($p_D$-value >0.8) items are sufficiently discriminating. The Cronbach's α value for the Differential Equations population is 0.822. Despite the high $p_D$-values (low difficulty) of many of the QuaRCS questions for this population, this statistic suggests that it may still be considered a valid assessment, albeit a generally easy one, for these students.

Despite the high average score, the Differential Equations population remains statistically distinct from the expert population, among whom none of the QuaRCS questions fall into the appropriate range for either difficulty **or** discrimination. This result demonstrates that the instrument is capable of measuring levels of quantitative ability among both general education introductory science students and more quantitatively advanced college students. It cannot distinguish levels of quantitative ability among experts, although it can identify them as a distinctly different population.

---

[16] The first and second University-required general education science courses.
[17] Note that these percentages do not sum to 100 because students are allowed to choose more than one major.

### Preliminary Pre/Post Analyses

During development of the QuaRCS, post-semester instruments were often administered in addition to pre-semester administrations. For the purposes of instrument development and validation, this practice allowed us to eliminate questions where discrimination declined significantly from pre- to post-assessment, as this trend is indicative of a low-quality item.

We consistently found that student scores did ***not*** increase over the course of the semester in any statistically significant way. For example, among students who "devoted effort" to the Fall 2013 instrument, the pre- ($N$=282, M=16.8, SD=4.3), mid- ($N$=175, M=17.6, SD=4.0) and post- ($N$=91, M=17.6, SD=4.4) score distributions are statistically indistinguishable.

Having shown that the instrument is capable of distinguishing various levels of quantitative ability, we can only conclude that students' skills are not improving as a result of taking these courses. ***This is unsurprising, as the QuaRCS development phase did not involve a study of curricular interventions focused on numeracy***. We developed this instrument precisely ***because*** we suspected that general education college science courses are not, as they are generally taught, serving to improve students' numeracy skills. The consistent lack of improvement in scores from pre- to post-semester assessments among students in our sample simply reinforces this assertion. Although some courses included in this study emphasized mathematics, this finding reinforces Steen and others' conclusions that "more mathematics does not necessarily lead to increased numeracy" (2001 p. 108). Whether such improvements are possible remains to be seen, and answering this question is the purpose of the broader QuaRCS study that is just now beginning.[18]

# Summary and Conclusions

This paper, the first in a series about the Quantitative Reasoning for College Science (QuaRCS) assessment, has focused on an overview of instrument development and validation that took place over the course of five years of administration in college classrooms. At the beginning of the paper, we highlighted the reasoning behind the development of a numeracy assessment instrument for college general education science courses and laid out three questions that we wish to address with the broader QuaRCS study. These are:

- Is it possible to make positive improvements in student numeracy skills or attitudes over the course of a single semester of college science?

- If quantitative skills are emphasized in a science course, are they then transferrable to "real life" contexts?

- How do students feel about the ways in which quantitative skills are emphasized in their courses? In particular, if they come into the course with a high level of ability, do they still benefit?

---

[18] We strongly encourage any educators engaged in innovative curricular interventions involving numeracy to consider administering the QuaRCS in their courses. Please contact us for more details.

The QuaRCS Assessment is focused on ten quantitative skills that were deemed important both for science literacy and for general educated citizenship by the 48 science and math educators completing our skills survey. Although math educators are more likely to label *all* numerical skills as important, science and numeracy educators generally agree on *which* numerical skills are most important. This agreement extends both to the importance of these skills in everyday life and to their importance for the understanding of science. The top five skills ranked most important in both contexts are: Graph Reading, Table Reading, Arithmetic, Proportional Reasoning, and Estimation.

We demonstrated that the QuaRCS is an effective assessment of numerical abilities and attitudes appropriate for the college general education science population. We described numerous test data collected and analyzed to assist in development and validation of the QuaRCS, and how these data helped to inform instrument refinements, including:

- The length of the survey (25 quantitative questions, 25 non-quantitative demographic and attitudinal questions) was shown to be free of the effects of test fatigue among students who expend effort.

- Question wording was extensively vetted through implementation of "free-response" format questions. These questions were used to generate authentic distractors, vet and clarify question phrasing, and to ensure that students could not arrive at a correct answer through incorrect reasoning.

- Both paper and online formats were explored. It was determined that the benefits of online assessment (out-of-class administration, ease of data collection, freedom from time constraints) outweighed the risks (lack of control for calculator and resource usage, idling).

Analysis of the data collected during development of the instrument further demonstrated that:

- The administration of confidence rankings after each quantitative question allows us to probe student awareness of numerical deficits. Preliminary results suggest that students are often "unconsciously incompetent."

- The addition of a question asking students to rank their effort on the assessment provided an important lens with which to view item and instrument statistics, particularly elapsed time and Cronbach's α.

- Cronbach's α has been consistently high throughout instrument development; it has improved as the instrument was refined, reaching 0.87 in the final version.

- Item discriminations are above 0.3 for the majority of items, and difficulties for all questions lie in the commonly accepted range for quality multiple-choice questions. Together, these indicate that the instrument is well matched to the general education science student population.

- Scores of experts and more advanced students are significantly different ($p<0.001$) than those of general education science students on the assessment, indicating that the instrument is capable of distinguishing the numerate from the innumerate.

- The instrument was also administered to a large Differential Equations course for STEM majors. Although many of the questions on the instrument are easier ($p_D>0.8$) for these students than generally considered acceptable for multiple-choice assessments, the discrimination values are still high, and Cronbach's α is well within the acceptable range for quality assessments.

- None of the post-semester score distributions accumulated during instrument development showed an improvement in score over their matching pre-semester assessment. This consistency argues for the urgency of the broader QuaRCS study, as it reinforces the assertion that general education college science courses as they are taught are generally ***not*** producing meaningful improvements to students' numerical skills. It also raises the question of whether such improvements are possible with innovative curricular techniques.

Having demonstrated the QuaRCS to be an appropriate and robust assessment for our purposes, we can now begin to use it to address some of the questions and concerns outlined at the beginning of this paper. Ultimately, we hope to answer definitively whether a semester of college science instruction is able to improve any one or more of the following: (a) students' transferable quantitative skills, (b) their attitude toward mathematics, or (c) their ability to recognize their own numeracy deficiencies. With a valid instrument in hand, we hope to identify exemplary instructors in this area and to inform the practices of all science educators in tackling the very important problem of innumeracy.

## Acknowledgments

This material is based in part upon work supported by the National Science Foundation under Grant Number 1140398. Any opinions, findings, and conclusions or recommendations expressed in this material are those of the author(s) and do not necessarily reflect the views of the National Science Foundation. We would like to thank Dr. Natasha Holmes, Dr. Carl Weiman, Dr. Heather Lehto, Dr. Len Vacher and the anonymous referees for their invaluable feedback in preparing this manuscript. We would also like to thank the members of Astro-Learner and the National Numeracy Network who participated in our expert surveys and analyses.

# Appendix

**Table A.1**
**Instrument Development Timeline**

| Semester | *N* Students Pre/[Mid]/Post (Matched) | Instrument Description | Summary of changes from previous version |
|---|---|---|---|
| Fall, 2010 | Instructor 1: 32 | **Format:** Scantron<br>**Administration:** In class. No credit<br>    Pre: Mid-semester<br>    Post: None<br>**Questions:**<br>  a)  3 Multiple Choice quantitative questions<br>1 Essay | N/A |
| Spring, 2011 | Instructor 1: 70/60 | **Format:** Scantron<br>**Administration:** In class. No credit<br>    Pre: First week of semester<br>    Post: Two weeks from end of semester<br>**Questions:**<br>  a)  8 Demographic + 5 attitude/skill self-assessment<br>  b)  10* Multiple choice quantitative questions<br>  c)  Confidence ranking after each quantitative question<br>  d)  3 questions reflecting on assessment<br>*22 total quantitative questions split into 3 versions, first 4 repeated for all students | (1)  Added demographic and attitude questions<br>(2)  Expanded quantitative question bank to 22<br>(3)  Added confidence ranking after EACH quantitative question<br>(4)  Pre AND post assessment |
| Fall, 2011 | Instructor 1: 156/163 (103)<br><br>Instructor 2: 22/22 (10)<br><br>Instructor 3: 29/6(5) | **Format:** Online<br>**Administration:** Out of Class. Participation credit.<br>    Pre: First two weeks of semester<br>    Post: Last two weeks of semester<br>**Questions:**<br>  (1)  11 demographic + 5 | (1)  Corrected for attrition and allow matched data by collecting names.<br>(2)  Several questions reworded for clarity based on focus group sessions<br>(3)  Encourage participation by assigning for participation credit (must offer alternate |
| | | | attitude/skill self-assessment<br>(2) 22 Multiple choice quantitative questions<br>(3) Confidence ranking after each quantitative question<br>(4) 3 questions reflecting on assessment | assignment per IRB)<br>(4) Recruited instructors request online format to save class time |
| Spring, 2012 | Instructor 1:<br>67/56 (40)<br><br>Instructor 2:<br>23/17 (11)<br><br>Instructor 4:<br>77/37 (22)<br><br>Instructor 5:<br>438/539 (278) | **Format:** Online<br>**Administration:** Out of Class. Participation credit.<br><br>Pre: First two weeks of semester<br>Post: Last two weeks of semester<br>**Questions:**<br>a) 12 demographic* + 5 attitude/skill self-assessment<br>b) 5 questions about course**<br>c) 22 Multiple choice quantitative questions<br>d) Confidence ranking after each quantitative question<br>e) 3 questions reflecting on assessment<br>*only name, course, age and instructor collected in post assessment<br>** post assessment only | (1) Added questions about course and whether and how quantitative skills were emphasized to post assessment<br>(2) Removed duplicate demographic question and added two questions about previous science coursework |
| Fall, 2013 | Instructor 1:<br>111/91/23 (21)<br><br>Instructor 6:<br>112/79/61 (50)<br><br>Instructor 7:<br>30/17/7 (4)<br><br>Instructor 8:<br>157/102/83 (58)<br><br>Instructor 9:<br>129/0/0 (0) | **Format:** Online<br>**Administration:** Out of Class. Participation credit.<br>Pre: First two weeks of semester<br>Mid: Week 11 of 16 week semester<br>Post: Last two weeks of semester<br>**Questions:**<br>a) Course, Instructor and Name<br>b) 25 Multiple choice quantitative questions<br>c) Confidence ranking after each quantitative question<br>d) 4 questions reflecting on assessment (incl. calculator usage)<br>e) 8 demographic + 8 attitude<br>f) Effort question | (1) To address question of whether late semester apathy is contributing to low post scores:<br>a. Added mid semester (post midterm) assessment<br>b. Added question asking students to quantify their effort on the assessment<br>(2) Major question rewordings and addition of table reading and area/volume skills<br>(3) Moved all demographic questions to end of assessment to mitigate stereotype threat<br>(4) Removed several demographic questions to make room for:<br>a. 5 Likert scale attitude questions<br>b. question about calculator usage on assessment<br>c. question about why chose major |
| Spring, 2014 | Instructor 2:<br>20/9 (6) | Same as Fall, 2013 except no mid-semester assessment | (1) Revised key math attitude question to Stapel scale[19] format, which forces students to place themselves on a scale between two opposite adjectives<br>(2) Addition of statistics as a skill category based on analysis of instructor surveys |

---

[19] Crespi (1961)

| Fall, 2014 | Instructor 6: 163 | **Format:** Online<br>**Administration:** Out of Class. Participation credit. Weeks 4-5<br>**Questions:**<br>  a) Course, Instructor and Name<br>  b) 25 Multiple choice quantitative questions<br>  c) Confidence ranking after each quantitative question<br>  d) 3 questions reflecting on assessment (incl. calculator usage)<br>  e) 8 demographic + 8 attitude<br>  f) Effort question | (1) Revised questions and distractors to reflect analysis of F13 and F14 free response administrations<br>(2) Added/revisited of several new questions to fill out skill categories, and removed several redundant or low-performing questions to make room<br>Added questions about ethnicity, disability status, and future coursework in mathematics |
| Spring 2012 | Instructor 1: $N$=30-72 | **Format:** Free Response Questions<br>**Administration:** In Class. Participation Credit. Throughout Semester.<br>**Questions:**<br>5 Open-Ended Free Response Versions of Questions from the Assessment | (1) Generated new authentic distractors based on misconceptions<br>(2) Reworded for clarity<br>(3) Revised in cases where students arrived at correct answer through incorrect reasoning |
| Fall, 2013 | Instructor 1: $N$=73-106 | **Format:** Free Response Questions<br>**Administration:** In Class. Participation Credit. Throughout Semester.<br>**Questions:**<br>11 Open-Ended Free Response Versions of Questions from the Assessment | (1) Generated new authentic distractors based on misconceptions<br>(2) Reworded for clarity<br>(3) Revised in cases where students arrived at correct answer through incorrect reasoning |
| Fall, 2014 | Instructor 1: $N$=35-68 | **Format:** Free Response Questions<br>**Administration:** In Class. Participation Credit. Throughout Semester.<br>**Questions:**<br>21 Open-Ended Free Response Versions of Questions from the Assessment | (1) Generated new authentic distractors based on misconceptions<br>(2) Reworded for clarity<br>(3) Revised in cases where students arrived at correct answer through incorrect reasoning |
| Spring, 2015 | | **FINAL VERSION** | |